\documentclass[a4paper,UKenglish,cleveref,autoref,thm-restate]{lipics-v2021}
    
\usepackage{booktabs}

\lstdefinelanguage{Solidity}{
  keywords={pragma, solidity, contract, mapping, address, uint, uint256,
            function, public, payable, view, internal, returns, require,
            if, throw, return, bool, true, false},
  keywordstyle=\bfseries,
  comment=[l]{//},
  commentstyle=\itshape,
  breaklines=true,
  numbers=left,
  numberstyle=\tiny,
  captionpos=t,
}
\theoremstyle{remark}
\newtheorem{finding}[theorem]{Finding}

\titlerunning{Pre-Model Representation Failures in GNN Detectors}

\title{Pre-Model Representation Failures in GNN-Based Smart Contract
       Vulnerability Detection}

\author{Birindwa Prisca Hondi}
        {Carnegie Mellon University Africa, Kigali, Rwanda}
        {pbirindw@andrew.cmu.edu}
        {https://orcid.org/0009-0001-9508-2744}
        {}

\author{Chinonso Philip Nwishienyi}
        {Carnegie Mellon University Africa, Kigali, Rwanda}
        {cnwishie@andrew.cmu.edu}
        {}
        {}

\author{Charity Wanja Mwaura}
        {Carnegie Mellon University Africa, Kigali, Rwanda}
        {cmwaura@andrew.cmu.edu}
        {}
        {}

\author{Alia Teto}
        {Carnegie Mellon University Africa, Kigali, Rwanda}
        {ateto@andrew.cmu.edu}
        {}
        {}

\author{Jema David Ndibwile}
        {Carnegie Mellon University Africa, Kigali, Rwanda}
        {jndibwil@andrew.cmu.edu}
        {https://orcid.org/0000-0002-7962-2237}
        {}

\authorrunning{B. Prisca Hondi, C. P. Nwishienyi, C. W. Mwaura, A. Teto, and J. David Ndibwile}

\Copyright{Prisca Birindwa Hondi, Chinonso Philip Nwishienyi, Charity Wanja Mwaura, Alia Teto, and Jema Ndibwile}



\nolinenumbers
\ArticleNo{35}
\begin{document}
\maketitle

\begin{abstract}
This paper is a failure analysis of the representation layer underlying
GNN-based smart contract vulnerability detectors. These systems convert
source code into graphs before any learning takes place; if the graph fails
to capture the code's semantics, no model improvement can compensate.

We investigate \textsc{GNNSCVulDetector} and identify four
failures. First, structurally different contracts produce byte-for-byte
identical graphs, constituting a concrete evasion attack. Second, graph
construction is governed by a hardcoded 47-entry variable whitelist (including
one duplicate entry), which constrains what the extractor can recognise. As a
consequence, identical vulnerabilities with different variable names produce
inconsistent graphs, graph quality degrades as naming diverges from the
whitelist, and when no entry matches the pipeline produces structural output
not grounded in source variables. Third, the~\texttt{C} node (the graph element
representing the external caller that triggers a reentrancy attack) is absent
from even the most canonical vulnerable contract in the literature. Fourth, a
controlled experiment confirms this as a direct misclassification: a fully
exploitable contract is labelled \emph{safe} because the $C \to W$ edge is
never constructed.

All four failures are demonstrated experimentally. Current accuracy figures
in the literature are measured under conditions that do not expose these
failures. We demonstrate one confirmed case of misclassification caused
directly by a representation-layer failure; the prevalence of such failures
in real-world contract populations remains an open empirical question.
\end{abstract}

\section{Introduction}
\label{sec:intro}

This paper is a failure analysis. Every experiment targets the representation
layer, the point at which source code is converted into a graph, to identify
failures that occur \emph{before} the model runs and that model improvement
alone cannot resolve.

Decentralized Finance (DeFi) has grown into a complex financial ecosystem built
on smart contracts. These contracts handle real assets and when they contain
vulnerabilities, the losses are immediate and irreversible. This has motivated
the development of automated detection systems, with Graph Neural Networks
(GNNs) emerging as a popular approach. Systems such as
DeFiGuard~\cite{defiguard}, DeFiTail~\cite{defitail},
GANS-MARL~\cite{gansmarl} and related systems~\cite{zhang2024, paramesha2024}
report detection accuracies between 88.9\% and 98.39\%.

These numbers look promising. But they all depend on something that has not
been adequately tested: the graph representation itself. Every GNN-based
detector works in two steps. First, source code is converted into a graph.
Second, the neural network analyses that graph. A GNN trained on millions of
contracts cannot detect a vulnerability whose defining features were silently
discarded during graph construction. No architectural improvement, no
additional training data and no fine-tuning can compensate for information
that was never extracted. This paper identifies a class of failures that model
improvement cannot address.

We investigated this through four experiments on
\textsc{GNNSCVulDetector}~\cite{zhuang2020}. Our findings are:

\begin{enumerate}
  \item \textbf{Structural invariance as an evasion attack.} Contracts with
    different names, different functions, and added dead code produce
    byte-for-byte identical graphs. This is a concrete evasion attack: an
    attacker who renames state variables bypasses detection with no knowledge
    of the model's training data or architecture.

  \item \textbf{A hardcoded 47-entry whitelist.} We designed four contracts
    with identical reentrancy vulnerabilities differing only in variable
    naming. Identical vulnerabilities produce inconsistent graph encodings,
    graph quality degrades progressively as naming deviates from the whitelist,
    and when no entry matches the pipeline produces structural output not
    grounded in source variables.

  \item \textbf{Missing caller context: an extraction inconsistency.}
    The \texttt{C} node (the graph element representing the external caller
    that triggers a reentrancy attack) is generated inconsistently: present
    for some contracts in the training data but absent for the most well-known
    vulnerable contract in the literature, with no signal from the pipeline
    that anything is missing.

  \item \textbf{Confirmed pre-model failure.} A controlled experiment on a
    purpose-built minimal contract shows the model outputs label~0
    (\emph{safe}) for a fully exploitable contract due to the absent
    $C \to W$ edge.
\end{enumerate}

These findings show that the pipeline fails at the representation layer before
the model runs. The accuracy figures reported in the literature reflect
performance under conditions that do not expose these failures.

\section{Threat Model}
\label{sec:threat}

We consider an attacker whose goal is to deploy a reentrancy-vulnerable smart
contract that evades GNN-based detection. The attacker has black-box access to
the detection pipeline: they know the system uses a GNN-based classifier but
have no knowledge of the model's weights, training data, or internal
architecture. The attacker's only capability is the ability to modify the
source code before deployment, specifically by renaming state variables,
functions, or the contract itself, and by inserting semantically inert dead
code. The reentrancy vulnerability itself, the
external-call-before-state-update pattern, is preserved intact.

Under this threat model, our experiments show that evasion is trivially
achievable: \cref{find:evasion} demonstrates that surface-level renaming alone produces
byte-for-byte identical graphs, and \cref{find:whitelist} shows that using any state
variable name outside the 47-entry whitelist produces a degenerate graph
regardless of the vulnerability present. Neither capability requires knowledge
of the model. The attacker needs only a text editor.

\section{Background and Related Work}
\label{sec:background}

\subsection{GNN-Based Smart Contract Detection}

GNN-based detectors convert smart contract source code into a graph and train
a neural network to recognise vulnerability patterns. Systems such as
DeFiGuard~\cite{defiguard}, DeFiTail~\cite{defitail}, and
GANS-MARL~\cite{gansmarl} report accuracies between 88.9\% and 98.39\% on
reentrancy and related exploits. \textsc{GNNSCVulDetector}~\cite{zhuang2020},
introduced at IJCAI~2020, converts Solidity source code into a graph and
classifies contracts using a GNN. This is the system we test.

\subsection{What Has Been Tested and What Has Not}

Existing evaluations test model accuracy on held-out contract sets.
GANS-MARL~\cite{gansmarl} also tests robustness against obfuscated inputs,
finding accuracy drops from 93.8\% to 88.3\%. However, all of these evaluations
assume the graph representation is correct. None examines whether the
representation preserves semantic information under modification. This paper
fills that gap.

\subsection{Adversarial Evasion Research}

Prior adversarial work focuses on bypassing the model through crafted inputs,
achieving evasion rates of 94--97\%~\cite{deleon2025,alharbi2025}. This paper
examines a more fundamental question: whether the representation given to the
model is faithful in the first place.

\section{Experimental Setup}
\label{sec:setup}

\subsection{Environment}

All experiments ran on a Kali Linux machine (x86\_64). GNNSCVulDetector was
originally built against an older TensorFlow/Keras stack, so we reproduced
that environment exactly, Python~3.7.17, TensorFlow~1.14.0, Keras~2.2.4,
scikit-learn~0.20.2 and protobuf~3.20.3, inside a dedicated virtual
environment to avoid dependency conflicts with newer libraries. Because
\textsc{GNNSCVulDetector} does not execute correctly under modern TensorFlow
releases, reproducing the original software environment was necessary to ensure
that observed failures originated from the published implementation rather than
from dependency incompatibilities introduced by version drift.

\subsection{Model and Pipeline}

\textsc{GNNSCVulDetector}~\cite{zhuang2020} works in three steps:
\begin{enumerate}
  \item \texttt{AutoExtractGraph.py} reads the Solidity source code and creates
    a graph of nodes and edges.
  \item \texttt{graph2vec.py} converts the graph into numerical vectors.
  \item \texttt{GNNSCModel.py} trains a GNN classifier on these vectors.
\end{enumerate}

Concretely, a graph here is a structured summary of the contract's logic: each
node represents a code element such as a state variable, a function, or an
external call, and each edge represents a relationship between them, for
example, ``this external call happens before this variable is updated.'' This
is conceptually similar to a flowchart of the contract's behaviour, except
built automatically from the source code rather than drawn by hand. The GNN in
step~3 never sees the original Solidity code; it only ever sees this graph. If
the graph omits or distorts a relationship, the model has no way to recover the
missing information, regardless of how well it is trained.

For Experiments~1 and~2, extracted graph files were compared directly at the
node, edge, node-feature, and edge-feature levels. Byte-for-byte identity
refers to exact equality of the serialised graph representations produced by
\texttt{AutoExtractGraph.py}. The representation is created in step~1. This
is where our investigation focuses.

\subsection{Dataset}

The dataset contains 6,200 labelled reentrancy contract graphs: 4,948 for
training and 1,252 for validation. Source contracts and pre-extracted graph
data are provided with the \textsc{GNNSCVulDetector}
repository~\cite{zhuang2020}.

\section{Baseline Performance}
\label{sec:baseline}

Table~\ref{tab:baseline} reports model performance when test contracts come
from the same distribution as the training data, that is, contracts using
variable names the extractor recognises and vulnerability patterns the graph
schema can represent. These figures represent the upper bound of performance
when the pipeline operates as designed; Sections~\ref{sec:exp1}--\ref{sec:exp4} test what happens when contracts fall outside these conditions. The threshold of 0.45 means the model classifies a contract as vulnerable whenever its output score exceeds 0.45; this value was set by the original authors and is used unchanged throughout our evaluation.

\begin{table}[tbp]
  \caption{Baseline performance on reentrancy detection (threshold = 0.45).
           Validation $\text{F}_1 = 0.848$ and AUC = 0.896 represent the
           performance ceiling when the pipeline operates as designed.
           False Positive Rate (FPR) is the share of safe contracts
           incorrectly flagged as vulnerable; AUC (area under ROC curve)
           summarises the model's ability to distinguish vulnerable from safe
           contracts across all thresholds, with 1.0 being perfect.}
  \label{tab:baseline}
  \centering
  \begin{tabular}{lcc}
    \toprule
    \textbf{Metric}      & \textbf{Training Set} & \textbf{Validation Set} \\
    \midrule
    F1-Score             & 0.933 & 0.848 \\
    Recall               & 0.926 & 0.865 \\
    Precision            & 0.940 & 0.805 \\
    False Positive Rate  & 0.028 & 0.109 \\
    Accuracy             & 0.955 & 0.882 \\
    AUC                  & ---   & 0.896 \\
    \bottomrule
  \end{tabular}
\end{table}

Each experiment that follows tests what happens when contracts fall outside
this distribution.

\section{Experiment 1: Structural Invariance as an Evasion Attack}
\label{sec:exp1}

\subsection{Setup}

We took \texttt{simple\_dao.sol}, a well-known reentrancy-vulnerable contract,
and created a modified version with the following changes:
\begin{itemize}
  \item Contract renamed: \texttt{SimpleDAO} $\to$ \texttt{VaultManager}
  \item Functions renamed: \texttt{donate} $\to$ \texttt{deposit},
    \texttt{withdraw} $\to$ \texttt{requestFunds},
    \texttt{queryCredit} $\to$ \texttt{checkBalance}
  \item State variable renamed: \texttt{credit} $\to$ \texttt{balances}
  \item Dead code inserted: internal function \texttt{\_processRequest} that
    performs no meaningful computation
\end{itemize}
The reentrancy vulnerability was kept intact. The external call still happens
before the state update.

\subsection{Original Contract}
Listing~\ref{lst:simpledao} shows the original vulnerable contract.
\begin{lstlisting}[caption={simple\_dao.sol: original vulnerable contract},
                   label={lst:simpledao}]
pragma solidity ^0.4.24;
contract SimpleDAO {
  mapping (address => uint) public credit;

  function donate(address to) payable public {
    credit[to] += msg.value;
  }

  function withdraw(uint amount) public {
    if (credit[msg.sender] < amount) { throw; }
    require(msg.sender.call.value(amount)()); // call FIRST
    credit[msg.sender] -= amount;             // state update AFTER
  }

  function queryCredit(address to) view public
      returns (uint) {
    return credit[to];
  }
}
\end{lstlisting}

\subsection{Modified Contract}
Listing~\ref{lst:simpledaomod} shows the modified version.
\begin{lstlisting}[float=tbp, caption={simple\_dao\_modified.sol: renamed and restructured,
                             vulnerability preserved},
                   label={lst:simpledaomod}]
pragma solidity ^0.4.24;
contract VaultManager {
  mapping (address => uint) public balances;

  function deposit(address to) payable public {
    balances[to] += msg.value;
  }

  function _processRequest(uint amount) internal
      returns (bool) {
    return amount > 0; // dead code
  }

  function requestFunds(uint amount) public {
    if (balances[msg.sender] < amount) { throw; }
    bool ok = _processRequest(amount);
    require(msg.sender.call.value(amount)()); // preserved
    balances[msg.sender] -= amount;
  }

  function checkBalance(address to) view public
      returns (uint) {
    return balances[to];
  }
}
\end{lstlisting}

\subsection{Results}

Both contracts were passed through \texttt{AutoExtractGraph.py}.
Table~\ref{tab:exp1} shows the output.

\begin{table}[tbp]
  \caption{Graph extraction output: original vs.\ modified. All features are
           identical across both contracts.}
  \label{tab:exp1}
  \centering
  \begin{tabular}{lccc}
    \toprule
    \textbf{Feature}  & \textbf{Original} & \textbf{Modified} & \textbf{Match?} \\
    \midrule
    Node count        & 5 & 5 & \checkmark \\
    Edge count        & 3 & 3 & \checkmark \\
    Node features     & Identical & Identical & \checkmark \\
    Edge features     & Identical & Identical & \checkmark \\
    C node present    & No & No & \checkmark \\
    \midrule
    \multicolumn{4}{c}{\textit{Graphs are byte-for-byte identical}} \\
    \bottomrule
  \end{tabular}
\end{table}

\begin{finding}[Evasion Attack]\label{find:evasion}
The graph extractor is invariant to contract renaming, variable renaming,
function renaming, and dead code insertion. The two contracts, clearly distinct programs, are assigned byte-for-byte identical graph representations. This is
not a theoretical edge case: it is a concrete evasion attack. An attacker who
renames state variables bypasses detection with no knowledge of the model's
training data or architecture. The rename alone is sufficient.
\end{finding}

\section{Experiment 2: The Variable Name Whitelist}
\label{sec:exp2}

\subsection{How the Extractor Constructs the Graph}

To understand why Experiment~1 produced identical graphs, we examined the
source code of \texttt{AutoExtractGraph.py}. The tool does not perform general
analysis of the Solidity code. Instead, it looks for specific variable names
using a hardcoded list, shown in Listing~\ref{lst:whitelist}:

\begin{lstlisting}[language=Python, float=tbp,
                   caption={Partial whitelist from \texttt{AutoExtractGraph.py}
                             (47 entries total, including one duplicate)},
                   label={lst:whitelist}]
var_list = [
  'balances[msg.sender]',
  'credit[msg.sender]',
  'credit[to]',
  'accountBalances[msg.sender]',
  'userBalance[msg.sender]',
  'Bal[msg.sender]',
  'payments[msg.sender]',
  'rewardsForA[recipient]',
  # ... 47 entries total
  'msg.sender'
]
\end{lstlisting}

If a match is found, a node is created in the graph. If no match is found, no
node is created. The duplicate entry is \texttt{participated[msg.sender]},
which appears at positions 2 and 23 of the list; its presence does not extend
coverage but does confirm the list was assembled manually rather than generated
systematically.

\subsection{Why Experiment 1 Produced Identical Graphs}

Both \texttt{credit[msg.sender]} and \texttt{balances[msg.sender]} appear on
the whitelist. Renaming the variable from \texttt{credit} to \texttt{balances}
moved it to another recognised entry. Both contracts produced the same graph
not because the tool understood they share the same vulnerability, but because
both names happen to be on the list.

\subsection{Experimental Proof of Whitelist Failure}

To empirically demonstrate the consequences of the whitelist mechanism, we
designed four contracts implementing identical reentrancy vulnerabilities that
differ only in the naming of the state variable. All four share the
external-call-before-state-update pattern. The four names were chosen to span
the whitelist boundary: Contract~A uses a name that matches the whitelist
exactly, Contract~B uses a name that partially overlaps a whitelist entry,
Contract~C uses a name with no \texttt{msg.sender} component and Contract~D
uses a name that matches no whitelist entry at all.

\begin{table}[tbp]
  \caption{Graph output across four identical vulnerabilities with different
           variable naming. Three distinct failure modes are evident:
           inconsistency (A vs.\ B), degradation (A vs.\ C), and fabrication
           (D).}
  \label{tab:exp2}
  \centering
  \begin{tabular}{clccp{3.8cm}}
    \toprule
    \textbf{Contract} & \textbf{Variable} & \textbf{Nodes} & \textbf{Edges}
      & \textbf{Observation} \\
    \midrule
    A & \texttt{balances[msg.sender]} (exact match)    & 5 & 3 & Normal graph \\
    B & \texttt{userDeposits[msg.sender]} (substring)  & 5 & 3 & Different edge type \\
    C & \texttt{ledger[owner]} (no msg.sender)         & 3 & 1 & Degraded graph \\
    D & \texttt{ledger[depositor]} (zero matches)      & 3 & 1 & Fabricated S node \\
    \bottomrule
  \end{tabular}
\end{table}

These results establish three empirical claims. Contracts~A and~B implement the
same vulnerability yet their graphs carry different edge types, so the model
receives structurally contradictory encodings for the same attack pattern based
solely on variable naming. Moving further from the whitelist, node count drops
from~5 to~3 and edge count from~3 to~1 between Contract~A and Contract~C, with
no indication from the pipeline that information has been lost. At the extreme,
Contract~D has zero whitelist matches yet the extractor still produces 3 nodes
and 1 edge: the~\texttt{S} node it emits does not correspond to any state
variable in the source; it is generated by a separate internal code path
triggered by the \texttt{.call.value()} pattern, producing structural output
not grounded in source variables.

\begin{finding}[Whitelist Failure]\label{find:whitelist}
The graph extractor relies on a hardcoded 47-entry variable name whitelist
(including one duplicate). Any contract whose state variable name falls outside
this list produces a degraded or fabricated graph with no warning from the
pipeline. Three failure modes are demonstrated: identical vulnerabilities
produce structurally inconsistent graphs when variable names differ
(Contracts~A and~B); graph quality degrades progressively as naming deviates
from the whitelist (Contract~C); and when no whitelist entry matches, the
pipeline fabricates structural output not grounded in any source variable
(Contract~D). This is a pre-model failure --- evasion occurs at the feature
extraction stage before any machine learning inference takes place.
\end{finding}

\section{Experiment 3: The Missing Caller Node---An Extraction Inconsistency}
\label{sec:exp3}

\subsection{What the C Node Represents}

In \textsc{GNNSCVulDetector}'s graph schema, the~\texttt{C} node represents
the caller function context: the external contract that initiates the call into
the vulnerable function. For reentrancy attacks, this is not optional context;
it is the attack's active agent. A reentrancy attack is defined by an external
contract calling the vulnerable function, which then makes an external call
back before updating its state. Without a~\texttt{C} node, the graph cannot
represent the attack mechanism at all.

The absence of the~\texttt{C} node is therefore not a minor limitation.
Table~\ref{tab:exp4} shows that \texttt{EtherStore.sol} in the training data
does receive a $C \to W$ edge, which confirms the schema is capable of encoding
the caller relationship under some conditions. The failure is one of
consistency: the extractor generates this edge for some contracts and silently
omits it for others implementing the same vulnerability, with no indication
from the pipeline that the representation is incomplete. This is a conceptual
failure in the extraction logic, not a strict impossibility in the schema.

\subsection{Results}

During graph extraction on both \texttt{simple\_dao.sol} and its modified
variant, the extractor emits:
\begin{quote}
\texttt{"There is no C node"}
\end{quote}
This occurs for the most well-known reentrancy-vulnerable contract in the
literature.

\begin{finding}[Extraction Inconsistency]\label{find:inconsistency}
Earlier versions of this work described this issue as a structural inability
to encode caller context. Our experiments instead show that the schema can
encode this relationship, but the extraction logic fails to do so consistently.
The $C \to W$ edge is generated inconsistently across contracts implementing
the same vulnerability. \texttt{EtherStore.sol} in the training data carries
this edge; \texttt{simple\_dao.sol} and its modified variant do not, despite
sharing the same reentrancy pattern. The extractor is therefore not structurally
incapable of encoding the caller relationship, but it fails to do so reliably.
The consequence is the same: the model receives graphs that omit the
cross-contract call dynamic that defines reentrancy for some contracts, with no
signal that anything is missing. The detector classifies based on incomplete
and inconsistently constructed representations of the attack it is trained to
find.
\end{finding}

\section{Experiment 4: Controlled Confirmation via userFunds.sol}
\label{sec:exp4}

This is the central empirical result of this paper.

\subsection{Motivation}

\cref{find:inconsistency} establishes that the~\texttt{C} node is absent for
\texttt{simple\_dao.sol}. To confirm that this absence directly causes a
misclassification, we constructed a minimal purpose-built contract
(\texttt{userFunds.sol}). We used \texttt{balances[msg.sender]} --- a variable
name that is on the whitelist --- so that any misclassification could not be
attributed to the whitelist failure described in \cref{find:whitelist}. This isolates a
single question: whether the missing $C \to W$ edge alone is sufficient to
cause the model to mislabel an exploitable contract as safe.
\texttt{userFunds.sol} was intentionally designed to isolate the caller-edge
failure while avoiding whitelist effects, by using
\texttt{balances[msg.sender]}, a variable name confirmed to be accepted by the extractor. Listing~\ref{lst:userfunds} shows the contract.

\begin{lstlisting}[caption={userFunds.sol: minimal controlled reentrancy
                             contract},
                   label={lst:userfunds}]
pragma solidity ^0.4.18;
contract UserFunds {
  mapping (address => uint256) public balances;

  function deposit() public payable {
    balances[msg.sender] += msg.value;
  }

  function withdrawFunds(uint256 _weiToWithdraw) public {
    require(balances[msg.sender] >= _weiToWithdraw);
    // VULNERABILITY: external call before state update
    msg.sender.call.value(_weiToWithdraw)();
    balances[msg.sender] -= _weiToWithdraw;
  }
}
\end{lstlisting}

\subsection{Graph Output and Comparison}

The~\texttt{C0} node was created but carries a \texttt{NULL} connection field
and appears in no edge; it is completely isolated. No $C \to W$ edge was
generated. An isolated node contributes nothing to the GNN's
message-passing computation: because graph neural networks learn by aggregating
information along edges, a node with no edges receives no neighbourhood signal
and passes none. The \texttt{C0} node exists in the graph data structure but
is invisible to the model's inference process.

\begin{table}[tbp]
  \caption{Graph structure comparison: \texttt{EtherStore.sol} vs.\
           \texttt{userFunds.sol}. The absent $C \to W$ edge in
           \texttt{userFunds.sol} directly produces the misclassification.}
  \label{tab:exp4}
  \centering
  \begin{tabular}{lcc}
    \toprule
    \textbf{Property}           & \textbf{EtherStore.sol} & \textbf{userFunds.sol} \\
    \midrule
    Ground truth label          & 1 (vulnerable) & 1 (vulnerable) \\
    C node present              & Yes            & Yes (isolated) \\
    $C \to W$ edge present      & Yes            & No             \\
    Graph encodes reentrancy    & Yes            & No             \\
    Model prediction            & Vulnerable \checkmark & Safe \texttimes \\
    \bottomrule
  \end{tabular}
\end{table}

\texttt{EtherStore.sol} from the training data carries edge \texttt{[1,7,2]}
encoding the $C \to W$ relationship. \texttt{userFunds.sol}, despite
implementing the same vulnerability, produces no such edge
(Table~\ref{tab:exp4}).

\subsection{Model Inference Results}

\begin{table}[tbp]
  \caption{\texttt{GNNSCModel.py} inference results, full 250-epoch run.}
  \label{tab:inference}
  \centering
  \begin{tabular}{lll}
    \toprule
    \textbf{Metric}           & \textbf{Value}  & \textbf{Notes} \\
    \midrule
    Total epochs run          & 250             & Full run completed \\
    Best training F1 epoch    & 242             & Train F1 = 0.9340 \\
    Best validation F1        & 0.8511          & Epoch 218 \\
    Final validation F1       & 0.8292          & Epoch 250 \\
    Final validation accuracy & 0.8786          & Epoch 250 \\
    Final validation recall   & 0.8601          & Epoch 250 \\
    Final validation FPR      & 0.1118          & Epoch 250 \\
    Predicted label (userFunds) & 0 (safe)      & Pre-model failure confirmed \\
    \bottomrule
  \end{tabular}
\end{table}

\begin{finding}\label{find:confirmed}
The model achieves a best validation $\text{F}_1$ of 0.8511 on the training
distribution yet outputs label~0 (\emph{safe}) for \texttt{userFunds.sol}, a
contract that is fully exploitable. The failure is located entirely in the
graph construction layer, prior to any model inference. The absent $C \to W$
edge means the GNN receives no structural signal encoding the reentrancy
relationship. The pipeline produces a graph, the model produces a label, and
the label is wrong, not because the model erred, but because the information
required to make the correct prediction was never provided to it.
\end{finding}

\section{Discussion}
\label{sec:discussion}

\subsection{Summary of Findings}

Our four experiments confirm that representation-layer failures are systematic
and occur upstream of the model. Because failures occur before the model runs,
downstream improvements cannot compensate.

\begin{itemize}
  \item Different contracts produce identical graphs, enabling evasion attacks
    (\cref{find:evasion}).
  \item The representation vocabulary is limited to 47 hardcoded entries
    (including one duplicate), producing inconsistent, degraded, and fabricated
    output (\cref{find:whitelist}).
  \item The most critical structural element of the targeted attack, the
    external caller, is absent from the graph schema. The $C \to W$ edge is
    generated inconsistently across contracts implementing the same
    vulnerability, meaning the model receives incomplete representations with
    no signal that anything is missing (\cref{find:inconsistency}).
  \item A fully exploitable contract is predicted as safe, directly caused by
    the absent $C \to W$ edge (\cref{find:confirmed}).
\end{itemize}

These are representation problems, not model problems. The neural network never
receives the information needed to make the correct decision.

\subsection{What This Means for Reported Accuracy}

Our results establish the existence of representation-layer failures and
provide one confirmed example in which such a failure leads directly to a
misclassification. In particular, \cref{find:confirmed} shows that a fully exploitable
contract can be labelled safe when the graph extraction process omits
information required to encode the reentrancy relationship.

At the same time, our study does not measure how frequently these failures
occur in real-world contract populations, nor does it estimate their aggregate
effect on detector accuracy. Quantifying that impact would require a
large-scale corpus study in which contracts are re-extracted and re-evaluated
under conditions designed to expose these failure modes.

The extent to which real-world contract populations fall outside the
extractor's 47-entry whitelist, and the resulting effect on detection
performance, therefore remains an open empirical question. Existing evaluation
protocols primarily measure model performance on supplied graph representations;
they do not explicitly test the representation-layer failure modes documented
in this paper.

\subsection{Scope: Why GNNSCVulDetector}

This study focuses on \textsc{GNNSCVulDetector}~\cite{zhuang2020} because it
is the foundational system in this line of work, widely cited as the basis for
subsequent GNN-based detectors, and provides full open-source access to both
the graph extraction pipeline and the trained model, making all findings
independently reproducible. The failures documented in this paper are
experimentally verified only for \textsc{GNNSCVulDetector}. Whether similar
vulnerabilities exist in other graph extraction pipelines remains an open
empirical question. We hypothesize that systems relying on comparable
surface-level matching mechanisms may exhibit related failure modes, but
evaluating this requires direct testing of those systems and is left for
future work.

\subsection{Generalizability as Future Work}

We therefore do not claim that the vulnerabilities identified here are
widespread. Rather, we present a reproducible methodology that can be applied
to additional GNN-based detectors to determine whether similar
representation-layer failures occur elsewhere. Applying the renaming-based
evaluation protocol used here to other systems would establish whether the
failure modes documented in this paper are specific to
\textsc{GNNSCVulDetector}'s implementation or reflect a broader pattern in
graph extraction design.

\subsection{Representation Before Model}

Most research in smart contract security focuses on improving the model: better
architectures, better training, more data. Our findings show that the more
urgent problem is upstream, since if the representation is broken, the model
cannot compensate. Future work should build representations that capture what
code does rather than what code looks like: graphs built from Abstract Syntax
Trees instead of name matching, control-flow analysis to detect the true order
of external calls and state updates, data-flow analysis to trace values
independent of naming, and explicit modelling of cross-contract calling
relationships so the external caller is represented rather than omitted.
Learned code embeddings such as CodeBERT are a plausible complement, replacing
a fixed vocabulary with representations that generalise across naming
conventions. Evaluation protocols must be extended accordingly, to cover
out-of-whitelist naming, structural transformation, and cross-contract
relationships.

These techniques share a common goal: representing program behaviour rather
than lexical appearance. An AST-based representation captures syntactic
structure independent of identifier naming, control-flow analysis captures
execution ordering, and data-flow analysis tracks value dependencies across
variables and functions. Together, these approaches reduce sensitivity to
superficial code transformations while preserving the semantic relationships
relevant to vulnerability detection.

\section{Conclusion}
\label{sec:conclusion}

This paper demonstrates that the graph representations used by GNN-based smart
contract vulnerability detectors do not faithfully capture semantic differences
between contracts. We show experimentally that structurally different contracts
produce identical graphs, that graph construction is governed by a 47-entry
whitelist that produces inconsistent, degraded, and fabricated output, that the
$C \to W$ edge encoding the external caller is generated inconsistently across
contracts meaning the graph schema fails to reliably represent the attack's
active agent, and that a fully exploitable contract is predicted as safe as a
direct consequence.

These failures are silent: the pipeline produces a graph, the model produces a
label, and no existing evaluation protocol checks whether the graph faithfully
represents the source contract. Existing evaluation protocols do not explicitly
evaluate the failure modes documented in this paper.

Closing this gap requires changes at the representation layer: moving from
surface-level pattern matching toward semantic analysis of what contracts do,
and evaluation protocols that expose the failure modes documented here.
Model-level improvements alone are insufficient.

\bibliographystyle{plainurl}
\bibliography{references}

\end{document}